\documentclass{article}
\usepackage{annals}
\title{The $XY$ Model and the Berezinskii-Kosterlitz-Thouless Phase Transition}
\name{Ralph Kenna (r.kenna@coventry.ac.uk)}
\address{Applied Mathematics Research Centre, Coventry University, Coventry, CV1 5FB, England}
\begin{document}
\maketitle
\begin{abstract}
In statistical physics, the $XY$ 
%or plane-rotator 
model in two dimensions provides 
the paradigmatic example of phase transitions mediated by topological defects (vortices). 
Over the years, a variety of analytical and numerical methods have been deployed in an
attempt to fully understand the nature of its transition, which is of the 
Berezinskii-Kosterlitz-Thouless type.
These met with only limited success until it was realized that subtle effects
(logarithmic corrections) that modify leading behaviour must be taken into account.
This realization prompted renewed activity in the field and significant progress has been made.
This paper contains a review of the importance of such subtleties, the role played
by vortices and of recent and current research in this area. 
Directions for desirable future research endeavours  
are outlined.
\end{abstract}

\section{Phase Transitions}
%%%%%%%%%%%%%%%%%%%%%%%%%%%%%%%%%%%%%%%%%%%%%%%%%%%%%%%%%%%%%%%%%%%%%%%%%%%%%%

Phase transitions are amongst the most remarkable and ubiquitous phenomena in nature.
They involve sudden changes in measurable macroscopic properties of systems and are brought 
about by varying external parameters such as temperature or pressure.
Familiar examples include the transitions from ice to water, water to steam, and 
the demagnetization of certain metals at  high temperature. 
These dramatic phenomena are described mathematically by non-analytic behaviour of 
thermodynamic functions, which reflect the drastic changes taking place in
the system at a microscopic level.
Besides materials science, phase transitions play vital roles in cosmology, particle 
physics, chemistry, biology, sociology and beyond;
The universe began in a symmetric manner and went through a series of 
phase transitions
through which the particles of matter with which we are familiar (electrons, 
protons, the Higgs boson, etc.) materialised. 
More mundane examples include traffic flow (where there is a transition between jammed
and free-flow\-ing states), growth phenomena and wealth accumulation 
(where there may be
a transition to a condensation phase, for example). While the latter examples
refer to non-equil\-ibrium systems, the emphasis in this article is on a famous
transition which exists in equilbrium systems.

The mathematical physics which describes such phenomena belongs to the realm of
equilibrium statistical mechanics, one of the most beautiful, sophisticated and successful
theories in physics. 
Equilibrium statistical physics is based on the following premise: the probability 
that a system is in a state $S$ with energy $E$ at a temperature $T$ is
\begin{equation}
 P(S) = \frac{e^{- \beta E (S) }}{Z_L(\beta)} \,,
\label{one}
\end{equation}
where  $\beta = 1/ k_B T$ and $k_B$ is a universal constant, and $Z_L(\beta)$ is a normalising factor known as
the partition function,
\begin{equation}
 Z_L(\beta) = \sum_S e^{- \beta E(S)}
\,.
\end{equation}
Here, the subscript $L$ indicates the linear extent of the system.

A related fundamental quantity is the free energy, $f_L(\beta)$, given by
\begin{equation}
 f_L(\beta) = \frac{1}{L^d}\ln{Z_L(\beta)}
\;,
\end{equation}
where $d$ is the dimensionality of the system.
Phase transitions can only occur when  the system under consideration has
an infinite number of states in which it can exist -- for example, 
in the thermodynamic limit $L \rightarrow \infty$.

In the modern classification scheme, such phase transitions are categorised as first-, second- (or higher-)
 order
if the lowest derivative of the free energy that displays non-analytic behaviour is the 
 first, second (or higher) one. 
Transitions of infinite order 
%are continuous and 
brake no system symmetries. 
The most famous of these is the Berezinskii-Kosterlitz-Thouless (BKT) transition in the
two-dimensional $XY$ model \cite{B,KT}.

\section{The two-dimensional $XY$ Model}
%%%%%%%%%%%%%%%%%%%%%%%%%%%%%%%%%%%%%%%%%%%%%%%%%%%%%%%%%%%%%%%%%%%%%%%%%%%%%%

The model is defined on a two-dimensional regular lattice, whose sites are labeled by the
index $i$, each of which is occupied by a spin or rotator $\vec{s}_i$.
These two-dimensional unit vectors have $O(2)$ or $U(1)$ symmetry. The energy of a given configuration
is
\begin{equation}
 E = - \sum_{\langle i, j \rangle }{\vec{s}_i\vec{s}_j}
\;, 
\label{XY}
\end{equation}
where the summation runs over nearest neighbouring sites or links.
This model is used to study systems such as films of superfluid helium, superconducting materials,
fluctuating surfaces, 
Josephson-junctions as well as certain magnetic, gaseous and liquid-crystal systems.

\begin{figure}[t]
\vspace{4cm}
\includegraphics{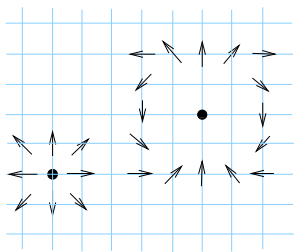}
\caption[a]{Examples of charge-$1$ (left) and charge-$2$ (right)
vortices in the $XY$ model on a lattice. Spins rotate
through $2\pi$ and $4\pi$, respectively, as contours around the
two bold points are traversed. }
\label{vortices}
\end{figure}
The scenario proposed in seminal papers by \mbox{Berezinskii} and Kosterlitz and Thouless \cite{B,KT} is that at a temperature
above the critical one ($T>T_c$ or $\beta < \beta_c$) positively and negatively charged vortices 
(i.e., vortices and antivortices) which are present (see Fig.~1) are unbound (dissociated from each other)
and disorder the system. 
Below the critical
temperature ($T < T_c$ or $\beta > \beta_c$) they are bound together and are relevant as dynamical degrees
of freedom.
There, long-range correlations between spins at sites $i$ and $j$ 
(separated by a distance $r$, say) exist and
are described by the correlation function whose leading behaviour 
in the thermodynamic infinite-volume limit is
\begin{equation}
 G_\infty(r) \sim r^{-\eta(\beta)}\,.
\end{equation}
The correlation length (which measures to what extent 
spins at different sites are correlated) diverges and 
this massless low-temperature phase persists, with the system remaining critical with
varying $\eta(\beta)$, up to $\beta = \beta_c$ at which $\eta(\beta_c) = \eta_c$.
Above this  point, correlations decay exponentially fast 
with leading behaviour
\begin{equation}
 G_\infty(r) \sim e^{-r/\xi_\infty(t)}
\,.
\end{equation}
Here $\xi_\infty(t)$ is the correlation length, and $t = T/T_c - 1 >0$ measures the distance from 
the critical point.

As this critical point is approached, the 
leading scaling behaviour of the correlation length, the specific heat and the susceptibility
(which respectively measure the response of the system to variations in the temperature 
and application of an external magnetic field) are
\begin{eqnarray}
 \xi_\infty (t) & \sim & e^{bt^{-\nu}} \; , 
\label{a}
\\
 C_\infty  (t)  & \sim & \xi_\infty^{-2} \; , 
\label{b}
\\
 \chi_\infty (t) & \sim & \xi_\infty^{2-\eta_c} \; ,
\label{c}
\end{eqnarray}
in which 
and $b$ is a non-universal constant.
This exponential  behavour is known as essential scaling, to distinguish
it from more conventional power-law scaling behaviour
(in which, for example, $\xi_\infty (t) \sim t^{-\nu}$). 
In summary, the BKT scenario means a transition which (i)
is mediated by vortex unbinding and (ii) exhibits essential scaling.

Besides the two-dimensional $XY$ model, transitions of the BKT type exist
in certain models with long-range interactions \cite{Ro97}, 
antiferromagnetic models \cite{CaVa98}, the ice-type $F$ model \cite{WeJa05}
and in string theory \cite{Ma02} amongst others. 
Thus a thorough and quantitative understanding 
of the paradigmatic $XY$ model is crucial to a wide breadth of theoretical physics.

For many years Monte Carlo and high-temperature analyses of the $XY$ model 
sought to verify the analytical BKT renormalisation-group (RG) prediction that $\nu = 1/2$
and $\eta_c = 1/4$ and to determine the value of $\beta_c$. Typically $\beta_c$ was determined by firstly fixing 
$\nu=1/2$. Subsequent measurements of $\eta_c$ yielded a value
incompatible with the the BKT prediction. 
Because of the elusiveness of its non-perturbative corroboration, the 
essential nature of the transition was  questioned 
\cite{Kim}. See Table~1 of \cite{IrKe97}
for an extensive overview of the status of the model up to that point.

\section{Logarithmic Corrections}
%%%%%%%%%%%%%%%%%%%%%%%%%%%%%%%%%%%%%%%%%%%%%%%%%%%%%%%%%%%%%%%%%%%%%%%%%%%%%%

The fundamental realization of \cite{IrKe96} was that the thermal scaling forms
(\ref{b}) and (\ref{c}) (and similar formulae for related thermodynamic functions) 
are inconsistent as they stand. Instead, they must 
be modified to include logarithmic corrections:
$ C_\infty (t) \sim  \xi_\infty(t)^{-2} \left( \ln{\xi_\infty(t)} \right)^{\tilde{q}}$
with  ${\tilde{q}}=6$ and 
\begin{equation}
\chi_\infty (t)  \sim \xi_\infty(t)^{2-\eta_c} \left( \ln{\xi_\infty(t)} \right)^{-2r} \; ,
\label{cc}
\end{equation}
where  RG  indications implicit in \cite{KT}
are that $r=-1/16 = -0.0625 $.

Eq.(\ref{cc}) is an analytic  prediction, which, since it is based on perturbation theory,
requires confirmation through 
non-perturbative approaches. However, infinite lattices are not 
attainable using finite numerical resources. Instead, one  simulates
finite systems, where, at criticality ($t=0$), the lattice size $L$ plays the
role of $\xi_\infty$ in this model. The finite-size scaling (FSS) prediction for the 
susceptibility is then 
\begin{equation}
\chi_L (0)  \sim L^{2-\eta_c} \left( \ln{L} \right)^{-2r} \; .
\label{ccFSS}
\end{equation}
To verify the BKT scaling scenario then,  the thermal formula (\ref{cc}) and/or the 
FSS formula (\ref{ccFSS}) needs to be confirmed numerically.
Sophisticated FSS techniques (involving partition function zeros)
were used in \cite{IrKe97,IrKe96} to resolve, for the first time, the hitherto conflicting
results for $\nu$,  $\beta_c$ and $\eta_c$. 
However, the analysis resulted in an estimate of $-0.02(1)$
for $r$, a value in conflict with the RG 
prediction of $r=-1/16 = -0.0625 $ from \cite{KT}.

Thus, in recent years, the focus of numerical 
studies of the $XY$ model shifted to the determination of the logarithm exponent $r$.
Indeed, the FSS analyses of \cite{Ja97,JaHa99,ChSt03}, using (\ref{ccFSS}),
yielded values compatible with 
that of \cite{IrKe97,IrKe96} but incompatible with \cite{KT} (see Table~1). 
Nonetheless, it was clear that taking the logarithmic corrections into account 
leads to the resolution of the $\nu$-$\beta_c$-$\eta_c$ controversy. 
The most precise estimate for the critical temperature  in the literature 
for the $XY$ model with the standard action of (\ref{XY}) is contained in \cite{HaPi97}
and is $\beta_c = 1.1199(1)$. This value was obtained by mapping the $XY$ model 
onto the exactly solvable body centered solid-on-solid model, 
and thereby circumventing the issue of logarithmic corrections.
As demonstrated in  Table~1, recent analyses which have inluded the logarithmic 
corrections have resulted in estimates for
$\beta_c$ compatible with this value (while keeping $\eta = 1/4$). 
In \cite{JaHa99} and \cite{ChSt03} respectively, 
phase transitions in a lattice grain boundary model and a lattice gauge theory are studied.
Because these are in the same universality class as the $XY$ model in two dimensions, 
they have the same scaling behaviour (\ref{ccFSS}). However, the critical temperatures 
in these models bear no relationship to that of the $XY$ counterpart.
Although \cite{Ja97} contains a study of the $XY$ model, it employs the Villain formulation, which is
different to (\ref{XY}), and leads to a different value for the nonuniversal quantity $\beta_c$.
\noindent
%\begin{table}[thp]
\begin{table}[th]
\caption{Estimates for $\beta_c$ and $r$ 
for the $XY$ model from a selection of recent
papers with indications of the method used to obtain them (RG related, FSS or thermal scaling).
}
\label{taba}
\begin{center}
\begin{small}
%\vspace{1.5cm}
%%%%%%%%%%%%%%%%%%%%%%%%%%%%%%%%%%%%%%%%%%%%%%%%%%%%%%%%%%%%%%%%%%%
\vspace{0.0cm}  
\noindent\begin{tabular}{|l|l|l|l|l|} 
    \hline 
    \hline
     Authors  &$\!\!\!$ Year  & $\!\!\!$ Method $\!\!\!\!$ & $\beta_c$ & $r$  \\
      & & & &  \\
    \hline
Kosterlitz,                           &$\!\!$1973& RG           &                        &  $\!\! -0.0625$        \\
   Thouless               \cite{KT}  &          &              &                        &                        \\ 
Irving,                               &$\!\!$1997& FSS          & $\!\!\!1.113(6)\!\!\!$ &  $\!\! -0.02(1)$       \\ 
     Kenna              \cite{IrKe97} &          &              &                        &                        \\ 
Patrascioiu,                          &$\!\!$1996& thermal$\!\!$&                        & $\!\!\!\quad 0.077(46)$\\
 Seiler                \cite{PaSe96}  &          &              &                        &                        \\ 
Campostrini                           &$\!\!$1996& thermal$\!\!$& $\!\!\!1.1158(6)\!\!\!$&  $\!\!\!\quad0.042(5)$ \\ 
    et al.                \cite{Ca96} &          &              & $\!\!\!1.120(4)\!\!\!$ &  $\!\!\!\quad0.05(2)$  \\
Janke                     \cite{Ja97} &$\!\!$1997& FSS          &                        & $\!\! -0.027(1)$       \\
                                      &          & thermal$\!\!$&                        &  $\!\!\!\quad0.0560(17) \!\!\!\!\!$ \\
Hasenbusch,                           &$\!\!$1997& RG           & $\!\!\!1.1199(1)\!\!\!\!\!\!\!\!\!\!\!\!\!\!\!$& \\
      Pinn              \cite{HaPi97} &          &              &                        &                        \\ 
Jaster,                               &$\!\!$1999& FSS          &                        & $\!\! -0.0233(10)\!\!\!$\\
 Hahn                   \cite{JaHa99} &          & thermal$\!\!$&                        &   $\!\!\!\quad0.056(9)$ \\
                                      &          &              &                        &  $\!\!\!\quad0.070(5)$  \\
Dukovski                              &$\!\!$2002& FSS          & $\!\!\!1.120(1)\!\!\!$ &                         \\
         et al.         \cite{DuMa02} &          &              &                        &                         \\
Chandrasekharan,$\!\!\!\!\!$          &$\!\!$2003& FSS          &                        & $\!\! -0.035(10)$       \\
       Strouthas        \cite{ChSt03} &          &              &                        &                         \\ 
Hasenbusch                \cite{Ha05} &$\!\!$2005& FSS          &                        & $\!\! -0.056(7)$        \\
 & & & & \\
    \hline
    \hline
  \end{tabular}
\end{small}
\end{center}
\end{table}

It was suggested in \cite{IrKe97,Ja97} that just as taking the logarithmic corrections into account 
leads to the resolution of the leading scaling behaviour,
in the same spirit it is conceivable that
numerical measurements for $r$ may become compatible with the RG prediction if
sub-leading corrections are taken into account. 
In this case, (\ref{ccFSS}) is more fully expressed as
\begin{equation}
\chi_L (0)  \sim L^{2-\eta_c} \left( \ln{L} \right)^{-2r} 
\left\{
 1 + {\cal{O}} \left( \frac{\ln{\ln{L}}}{\ln{L}} \right)
\right\}
\; .
\label{ccFSSc}
\end{equation}
This was tested in \cite{IrKe97,Ja97} to determine if the discrepancies between the numerical 
and theoretical estimates for $r$ can be ascribed to sub-leading corrections.
However, the lattice sizes available (up to $L=256$) were too small to  resolve the issue.

Recently this issue was again addressed in a very high precision numerical simulation 
(using lattices as big as $L=2048$) in \cite{Ha05}. 
Using the ansatz 
\begin{equation}
\chi_L (0)  \sim L^{2-\eta_c} \left( C +  \ln{L} \right)^{-2r} 
\; ,
\label{ccFSSH}
\end{equation}
and fitting to $r$ gives $r=-0.056(7)$ if large enough 
lattices are used. This value is compatible with the analytic prediction, $r=-0.0625$.

Notwithstanding this result, which is FSS based, it is 
rather surprising that all of the analyses of the thermal scaling formula (\ref{cc}) in Table~1  
yield positive values of $r$ 
far from the RG prediction
that $r = -0.0625$. It appears, therefore, that the FSS approach is more powerful than that
based on thermal scaling, although more extensive analyses  would
be required to test how this approach compares with  that of \cite{Ha05}.

While large scale simulations were required to resolve these
puzzles, 
%(at least within the FSS approach), 
a different technique, not requiring extensive simulations
was used in \cite{Berche}.
A conformal mapping was used to switch from a confined $L \times L$ lattice to the semi-infinite half plane.
In this way, $\eta(\beta)$ could be deduced from the correlation function at any value of 
$\beta \ge \beta_c$. 
% from relatively small lattices.
In particular, $\eta_c=1/4$ is accurately recovered at the transition temperature and 
clear evidence for  existence of logarithmic corrections in the correlation function $G(r)$ is presented.
%A precise determination of $r$  was not possible in \cite{Berche} (since the lattices are still too small).

\section{Vortex Unbinding}
%%%%%%%%%%%%%%%%%%%%%%%%%%%%%%%%%%%%%%%%%%%%%%%%%%%%%%%%%%%%%%%%%%%%%%%%%%%%%%

The vortex-binding scenario is crucial to the BKT phase transition in the 
two-dimensional $XY$ model. 
Because the energy of a single vortex increases with the system size as $\ln{L}$, 
at low temperature they can only occur in  vortex-antivortex pairs. 
Mutual cancelation of their individual ordering effects means that such 
a pair can only affect nearby spins and 
cannot significantly disorder the whole system. Topological long-range order exists in the system 
at low temperature.
However at high temperature, the number of vortices proliferates and the distance between erstwhile
partners becomes so large that they are effectively free and render the system disordered.

%This vortex-binding mechanism was examined in \cite{IrKe97}. In particular, i

It was for a long time believed that altering the energetics of the $XY$ model  to
disable the vortex-binding scenario may lead to a transition different to the 
Berezinskii-Kosterlitz-Thouless one \cite{SaWi88}.
The step model is obtained from the $XY$ model by replacing the  Hamiltonian (\ref{XY}) 
by 
\begin{equation}
 E = - \sum_{\langle i, j \rangle }{{\rm{sgn}}\left( \vec{s}_i\vec{s}_j \right) }
\;. 
\label{step}
\end{equation}
The energy associated with a single vortex for this system is 
expected to be independent of the lattice size. Therefore, on the basis of this argument,
vortices could  exist at all temperatures, disordering the system. 
I.e., there was expected to be no vortex-driven phase transition in the step model -- 
if there is a phase transition, it was expected not to be of the BKT type.
Indeed, early studies supported this assertion \cite{SaWi88}.

However, in \cite{IrKe97,IrKe96}, very strong numerical evidence was presented that
(i) there is a phase transition in the step model and (ii) it is of the BKT type 
(with even the corrections to scaling being the same as those for the $XY$ model).
Given the very different vortex energetics for the two models, this came as a surprise.

The issue was further addressed in \cite{OlHo01}, where evidence for the existence
of a BKT transition in the step model was again proffered. 
The approach of \cite{OlHo01} focused on numerical analyses of the helicity modulus, 
which experiences a jump at the transition. 
A similar approach to the $XY$ model is contained in \cite{MiKi03}.
The main idea of \cite{OlHo01}, which explains the occurence of the BKT transition in the
step model, is that while the energy associated with a single fixed vortex in the system remains finite, 
the {\emph{free energy}} grows as $\ln{L}$.
This fact inhibits proliferation of free vortices in the low-temperature phase.
It further implies that the harmonic properties of the interaction (\ref{XY}) do
not form a necessary condition for a BKT transition. 
Consequently, and as pointed out in  \cite{OlHo01}, 
the BKT  phase transition may be an even more general phenomenon than hitherto
recognised.

\section{Future Directions}
%%%%%%%%%%%%%%%%%%%%%%%%%%%%%%%%%%%%%%%%%%%%%%%%%%%%%%%%%%%%%%%%%%%%%%%%%%%%%%

{\bf{Asymptotic freedom of $d=2$  $O(N)$ models:}}
It is generally believed that there are differences of a fundamental nature between 
abelian and nonabelian models. 
The $XY$ model has an $O(2)$ symmetry group and is abelian,
while all $O(N)$ models with $N > 2$ are nonabelian.
The Mermin-Wagner theorem states that 
a continuous symmetry of the $O(N)$ type cannot be broken in two dimensions \cite{MW}.
Thus there cannot be a transition to a phase with long-range order
in either of the $N=2$ or $N>2$ scenarios there. 
However, in a two-dimensional theory, topological defects of dimension $m$ can exist 
if the $(1-m)^{\rm{th}}$ homotopy group, $\pi_{1-m}$, of the order parameter space 
is non-trivial. 
For $O(N)$ models, this space is the hypersphere $S^{N-1}$.
The only non-trivial group is $\pi_{1}(S^{1})$ which is isomorphic to the set of integers under addition.
This is the condition that gives rise to point defects (vortices) 
with integer charge in the  $N=2$ case (the $XY$ model).
The binding of these vortices at low temperature is the mechanism giving rise to the BKT phase transition 
\cite{B,KT}.

For $N>2$, conditions are not supportive of the existence of topological defects of this type and
the widely held belief is that there is no phase transition, there being no distinct low-temperature phase. 
Perturbation theory predicts that the $N>2$  models are asymptotically free.
There is, however, no rigorous proof to this effect.
This belief has been  questioned and \cite{PaSe96,PaSc92}
have given numerical evidence for the existence of phase transitions of the BKT type in these models
as well as
heuristic explanations of why such transition could occur
and a rigorous proof that this would be incompatible with asymptotic freedom.

Perturbative and Monte Carlo calculations for the $N=3$ and $N=8$ models have been performed 
in \cite{AlBu99}, which do not support the existence of such a BKT-like phase transition there
and instead  are in agreement with perturbation theory and the asymptotic freedom scenario.
Nonetheless, the controversy has not entirely gone away \cite{Ba01,Pa01,Se03} , and one may argue that 
inclusion of logarithmic considerations could help for a precise unambiguous resolution.

~\\
{\bf{The diluted $XY$ model:}}
An interesting current topic of research has been the question of the role of impurities in the $XY$ model \cite{Ro99}.
The presence of impurities brings the model closer to real systems, where such physical defects are present. 
Impurities are modeled by randomly diluting the number of sites (or bonds) on the lattice.
Clearly, if the dilution is so strong as to inhibit the percolation of spin-spin interactions
across the lattice (such that it is effectively broken into finite disconnected sets) no phase transition can occur
for any model.
Thus  moderate dilution generally is expected to decrease the location of the transition temperature.
However, the special additional
feature of the $XY$ model is the presence of vortices and the fact that they drive the transition.
Vortices are attracted to and, to some extent, anchored by impurities and the vortex energy
is reduced at such a vacancy. 
Therefore with increasing dilution, more vortices can be formed and the amount of
disorder in the system is increased. 
This effect may enhance the lowering of the critical temperature to such an extent that 
it vanishes before the percolation threshold is reached. 

This is the issue addressed in \cite{Ro99,LeCo03,BeFa03,WyPe05}.
The percolation threshold occurs when the density of site vacancies is $\rho \approx 0.41$.
The critical vacancy density  is identified by the vanishing of 
the critical temperature, 
which in turn is identified as being the location at which $\eta(\beta)=1/4$.
In \cite{LeCo03}  the critical temperature was reported to vanish above the percolation threshold
at vacancy density $\rho \approx 0.3$.
However, in \cite{{BeFa03}} it was suggested that the critical density 
is, in fact, closer to the percolation threshold. 
Support for the latter result recently appeared in  \cite{WyPe05}.
If this is true, it means that the vortices do not, in fact, strongly enhance the lowering of the 
critical temperature.

From the collective experience with the pure $XY$ model as reported above, it is clear that
a more precise identification of the critical temperature through $\eta(\beta)=1/4$ 
would require taking account of 
the logarithmic corrections, although ignoring them may suffice as first approximation.
Indeed, ignoring these corrections in the pure model 
also leads to an estimate for the critical temperature
 which is higher than accurate values \cite{WyPe05}.

Besides the value of the critical temperature in a diluted model, one is also interested
in the scaling behaviour of the thermodynamic functions at the phase transition there.
The Harris criterion predicts that disorder does not change the leading scaling behaviour of a model
if the critical exponent $\alpha$ associated with the  specific heat of the pure model 
is negative \cite{Ha74}. 
This is the case for the $XY$ model in two dimensions.
However, it is unclear what effect dilution can have on the quantitative nature of the
exponents of the logarithmic corrections in such a case. 
This would be an interesting avenue for future research, and the $XY$ model 
(which has negative $\alpha$)
offers an ideal platform upon which to base such pursuits.

~\\
{\bf{Other models with logarithmic corrections:}}
Logarithmic corrections to scaling also exist in other important models.
While their existence has been unambiguously verified in $d=4$ $O(N)$ models \cite{KeLa94,Ke04},
this is not so in most cases \cite{us}. 
In particular, Kim and Landau applied FSS techniques to the four-state $d=2$ Potts model \cite{KiLa98}.
Again, FSS behaviour could only be described if the logarithmic corrections are included. 
However, here inclusion of the leading multiplicative logarithmic corrections is insufficient
and sub-leading additive corrections are required.
The puzzle of why unusually large numbers of correction terms are necessary (despite the 
availability of the exact value of $\beta_c$ in this model) could, perhaps, now be resolved 
by an analysis on the scale of \cite{Ha05} and this would be another 
interesting avenue to pursue.
Similar problems concerning logarithmic corrections and their detection 
exist in other two-dimensional
models such as  the diluted Ising model \cite{RoAd99}.
Theoretical progress on the general issue of logarithmic corrections  will be reported elsewhere \cite{us}.

\section{Conclusions}
%%%%%%%%%%%%%%%%%%%%%%%%%%%%%%%%%%%%%%%%%%%%%%%%%%%%%%%%%%%%%%%%%%%%%%%%%%%%%%

The issue of  topologically driven phase transitions 
characterized by the two-dimensional $XY$ model has been revisited 
and a timely review of the status of  scaling at the famous Berezinskii-Kosterlitz-Thouless
transition  given. 
After two dec\-ades of work, the perturbative renormalization group precictions 
of \cite{B,KT} for the leading scaling behaviour were confirmed in \cite{IrKe97,IrKe96}
and there is now little or no doubt about the correctness of the analytic predictions, $\nu = 1/2$
and $\eta_c = 1/4$. 
The recent controversy over the value of the leading logarithmic correction exponent $r$ 
(summarized in Table~1)
has been re-examined and claims as to its resolution (at least in the context of finite-size scaling)
and state-of-the-art calculations summarized \cite{Ha05}. 
Besides the $XY$ model, such multiplicative logarithmic corrections are manifest in
a variety of other models, and their resolution in these contexts is 
now at the forefront current modern numerical investigations in statistical and lattice physics.
A summary of some recent work on such models, as well as likely future directions
has been given.

Finally, recent work \cite{OlHo01} confirming the vortex-binding scenario as the phase transition mechanism 
in the $XY$ model  has  also been reviewed
to complete a full account of the current status of one of the most remarkable and beautiful 
models in theoretical physics.

\section{Acknowledgements}
%%%%%%%%%%%%%%%%%%%%%%%%%%%%%%%%%%%%%%%%%%%%%%%%%%%%%%%%%%%%%%%%%%%%%%%%%%%%%%

This work was supported by EU Marie Curie Research Grants 
CHBICT941125 and FMBICT961757.

\end{document}